\documentclass[aps,twocolumn,pre,superscriptaddress,notitlepage]{revtex4-2}
\usepackage{tikz, bm}
\usetikzlibrary{arrows.meta,arrows}
\usepackage{bm,amsmath,amssymb,graphicx, mathtools,multirow,gensymb}
\usepackage[colorlinks=true,linkcolor=MidnightBlue,urlcolor=black,citecolor=MidnightBlue,anchorcolor=MidnightBlue]{hyperref}\usepackage[dvipsnames]{xcolor}

\newcommand{\tpix}{\ensuremath{\frac{2e^{-q^2}}{\sqrt{\pi}}}}

\begin{document}
\title{Ewald summing irreducible components of flow around active particles}
\author{Mayurakshi Deb}
\affiliation{Department of Physics, Indian Institute of Technology Madras, Chennai, India}
\author{Rajesh Singh}
\email{rsingh@physics.iitm.ac.in}
\affiliation{Department of Physics, Indian Institute of Technology Madras, Chennai, India}
\begin{abstract}
We present a method to compute Ewald summation for the irreducible components of flow around active particles to study hydrodynamic interactions in active colloidal suspensions.
An active particle is modeled as a colloidal sphere with a surface slip velocity.
Using this model, we obtain an irreducible representation of the fluid flow produced by an active particle in periodic geometry of Stokes flow for an arbitrary surface slip. The solution of the active
flow is obtained in terms of lattice sum of the Oseen tensor and their derivatives. The lattice sum
is accelerated using the Ewald summation technique.
We apply the method to compute explicit expression for rigid body motion of hydrodynamically interacting active particles. Our method presents a way for dynamic simulation of active particles due to arbitrary mode of active slip in periodic geometry of Stokes flow. 
\end{abstract}
\maketitle

\section{Introduction}

Suspensions of active colloidal particles, on whose surfaces non-equilibrium
process of chemical (autophoretic colloids \cite{paxton2004,howse2007self,jiang2010active,palacci2013living})
or biological origin (microorganisms \cite{drescher2009,drescher2011,petroff2015fast,brennen1977,ebbens2010pursuit})
takes place, have been the subject of much recent study \cite{shaebani2020computational}. The thickness
of the non-equilibrium boundary layer is small compared to the colloidal
size, and thus, the effect of activity appears as an augmentation
of the usual no-slip boundary condition by an active slip \cite{anderson1989colloid}.
The slip drives exterior flow and resulting fluid stress may lead to the
self-propulsion of the colloid (even in absence of an external body force on the colloid).
The exterior flow created by colloidal particles leads to hydrodynamic interactions between
the colloids. 

Recently, a theory has been developed to study hydrodynamic interactions of active particles by modeling them as spheres with surface slip in unbounded 
\cite{singh2015many,singh2018generalized,turk2022stokes} 
and wall-bounded domain \cite{singh2016crystallization,turk2024fluctuating}. 
In this work, we extend
the theory to compute hydrodynamic interaction of $N$ spherical
particles with active boundary layers in periodic geometries of Stokes flow. 
Our explicit expressions from Ewald summing irreducible components of flow around active particles presents a way for simulations of active particles in periodic domain due to arbitrary mode of active slip. 
In this paper, we present a formal expression for the Ewald summation for an arbitrary mode of the active slip and provide explicit forms for the modes up to $l=4$ of the active slip. Thus, we cover all long-ranged modes which decay as $1/r^3$ or slower, where $r$ is the distance of the field point from the source point. In addition, we give explicit expression of the flow due to $l\sigma=4a$ mode of the slip which decays as $1/r^4$. This mode is lowest mode capable of producing an active spinning of an isolated active particle \cite{ghose2014irreducible,pak2014generalized,pedley2016squirmers}. Thus, this mode is useful in modeling a suspension of self-spinning green algae \emph{Volvox} \cite{drescher2009,goldstein2015green}. To the best of our knowledge, Ewald sum of all these higher irreducible modes (beyond $l>2$) is not available in the literature. The explicit forms for Ewald summed expression for Stokes flow around active colloidal particles are only available for Stokeslet ($l\sigma=1s$), stresslet ($l\sigma=2s$), and rotlet ($l\sigma=2a$) \cite{beenakker1986ewald, bagge2023fast,adhyapak2018ewald}.

The remainder of the paper is organized as follows. In section \ref{MAC},
we present expressions
of irreducible components of flow around active particles. The equations describing the rigid body motion of active colloids is given in section \ref{sec:RBM}. These expressions have been summed using
Ewald technique for accelerated convergence in numerical simulations with periodic boundary condition. Explicit forms of these expressions are provided for direct numerical simulations of active suspensions. 
{
In section \ref{sec:app}, we compare the full Ewald summation technique to the minimum image convention for explaining the need of Ewald summation for higher modes of activity.}
Finally, we conclude in section \ref{sec:Conclusion} by summarizing our results 
and suggesting
future directions. 
In what follows, we detail our results.

\section{Irreducible components of flow around  active particles\label{MAC}}

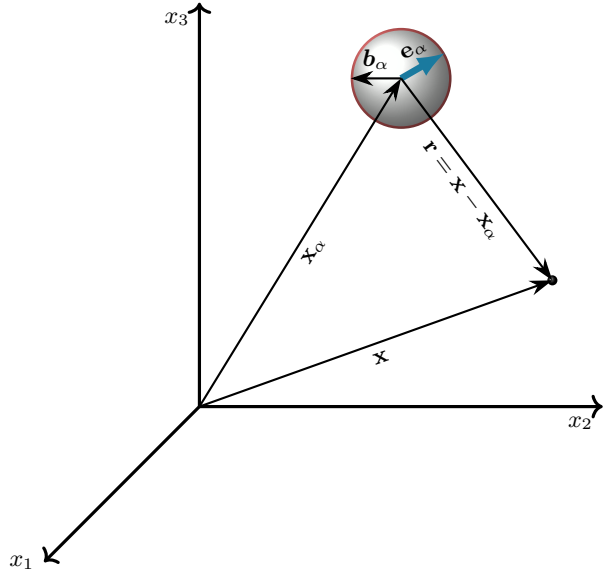
\begin{figure}
\centering
\begin{tikzpicture}[scale=0.667]
\node (s1) at (4, 6.5) {};   
\node (s2) at (7, 2.5) {};   
\draw[very thick,->] (0,0,0) -- (0, 0, 8) node[anchor=east]     {$x_1$};
\draw[very thick,->] (0,0,0) -- (8, 0, 0) node[anchor=north east]{$ x_2$};
\draw[very thick,->] (0,0,0) -- (0, 8, 0) node[anchor=north east]{$x_3$};
\shade [ball color=red!56!] (s1) circle [radius=1.0cm];         
\shade [ball color=blue!10!green!1!] (s1) circle [radius=0.95cm];              
\shade [ball color=black!256!] (s2) circle [radius=0.1cm];         
\draw [thick,-{Stealth[length=3mm, width=2mm]}] (0, 0)--(4, 6.5) node [midway, below, sloped] {{$\mathbf{x}_\alpha$}};
\draw[->, >=latex, green!42!blue!90, line width=2.5pt] (4, 6.5)--(4.866, 6.9999) node[midway, above, black, sloped]{$\mathbf{e}_\alpha$} ;
\draw [thick,-{Stealth[length=3mm, width=2mm]}] (4, 6.5)--(3, 6.5) node[midway, above, black, sloped]{$\bm{b}_\alpha$} ;
\draw [thick,-{Stealth[length=3mm, width=2mm]}] (4, 6.5)--(7,2.5) node [midway, below, sloped] {{$\mathbf{r}=\mathbf{x} - \mathbf{x}_\alpha$}};
\draw [thick,-{Stealth[length=3mm, width=2mm]}] (0, 0)--(7,2.5) node [midway, below, sloped] {{$\mathbf{x}$}};
\end{tikzpicture}
\caption{
    Coordinate system for the $\alpha$-th active particle, which is a sphere centered at  ${\mathbf{x}}_\alpha$.
    { The radius vector $\bm{b}_\alpha$ is drawn from the center of the $\alpha$th particle to a point on its surface. 
    On the other hand, $\mathbf{e}_\alpha$ is the orientation of the $\alpha$th active particle (it points in the direction of self-propulsion of an isolated active particle).}
    The field point is denoted by the position vector ${\mathbf{x}}$. The vector  ${\mathbf{r}} = {\mathbf{x}} - \mathbf{x}_\alpha$ shows the displacement between the field and source points.
} 
\label{coordinate-system}
\end{figure}

We consider a system of $N$ active colloidal spheres of radius $b$
suspended in an incompressible viscous fluid of dynamic viscosity
$\eta$ and volume $\mathcal{V}$. The $\alpha$th sphere centered at $\mathbf{x}_{\alpha}$ has radius
vector $\boldsymbol{ b }_{\alpha}$, while its orientation 
is given by the unit vector $\mathbf{e}_{\alpha}$. See Fig.\ref{coordinate-system}. {An isolated active particle self-propels along its orientation vector}.
We denote the translational velocity and angular velocity of the sphere as $\mathbf{U}_{\alpha}$
and $\boldsymbol{\Omega}_{\alpha}$ respectively. We model active colloids
as spherical colloid with a surface slip $\boldsymbol{v}^{\mathcal{A}}$ \cite{anderson1989colloid,ebbens2010pursuit}. 
Fluid velocity on the surface of active particle is:
\begin{equation}
    \boldsymbol{v}(\mathbf{x}_\alpha+\boldsymbol{b}_\alpha)=\mathbf{U}_\alpha+\boldsymbol{\Omega}_\alpha\times\boldsymbol{b}_\alpha+\boldsymbol{v}^{\mathcal A}_\alpha(\boldsymbol{b}_\alpha).
    \label{eq:slip}
\end{equation}
The slip velocity $\boldsymbol{v}_\alpha^{\mathcal A}$ arises from interfacial forces. This could be due  motion in ciliary layer of a microorganism \cite{lighthill1952,blake1971a,brennen1977} or due to gradients in chemical potential or temperature on the surface of a colloidal particle \cite{anderson1989colloid, ebbens2010pursuit}. 
The active slip can be of arbitrary functional form as long as it satisfies the restriction  that it
conserves mass in the fluid, i.e. $\int \boldsymbol{v}^{\mathcal A}_\alpha(\boldsymbol{b}_\alpha)\cdot \hat{\boldsymbol b}_\alpha\,\mathrm{d}\mathcal S_\alpha=0$. Here $\hat{\boldsymbol b}_\alpha = \boldsymbol{b}/b$ is a unit vector along the radius of the $\alpha$-th particle, while $b=|\boldsymbol{b}|$. In this paper, we use
Greek alphabets - such as $\alpha,\beta$ - to denote particle indices, while Cartesian indices are written using {Roman alphabets} $i,j$, etc.

At the colloidal scale, the fluid flow $\boldsymbol{v}$ satisfies
the Stokes equation \cite{pozrikidis1992}:
\begin{subequations}
\begin{gather}
\nabla_i v_i=0,\quad
\nabla_i {\sigma}_{ij}=0,\\
 {\sigma}_{ij}=-p\,\delta_{ij}+\eta
\left[ 
{\nabla}_i {v}_j+ {\nabla}_j{v}_i
\right].
\end{gather}
\end{subequations}Here ${\sigma}_{ij}$ is the stress in a Newtonian fluid,
$p$ is fluid pressure, while $\delta_{ij}$ is the Kronecker delta. 
We use Einstein summation convention over repeated Cartesian indices ($i,j,\dots$), unless specified otherwise.

\subsection{Boundary integral formulation}\label{sec:BIE}
The flow velocity $\boldsymbol v(\mathbf{x})$, at any point $\mathbf{x}$ in the bulk fluid ($\mathbf{x} \in\mathcal V$) due to particles located at $\mathbf{x}_\alpha$ (see Fig.\ref{coordinate-system}), is obtained from the boundary integral representation
of Stokes flow \cite{fkg1930bandwertaufgaben,ladyzhenskaya1969,youngren1975stokes,zick1982stokes,pozrikidis1992,muldowney1995spectral,cheng2005heritage,leal2007advanced,masoud2019reciprocal,singh2015many}
\begin{align}
v_{i}(\mathbf{x})&=- \sum_{\alpha=1}^N\int_{\mathcal{S}_\alpha} G_{ij}(\mathbf{x},{\mathbf{r}_{\alpha}})f_{j}(\mathbf{r}_\alpha)\,\mathrm{d}\mathcal{S}_\alpha
\nonumber\\
&+ \sum_{\alpha=1}^N\int_{\mathcal{S}_\alpha} K_{jik}({\mathbf{r}_{\alpha}},\mathbf{x})\,n_{k}\,v_{j}(\mathbf{r}_\alpha)\,\mathrm{d}\mathcal{S}_\alpha.
%
\label{eq:BIE-PBC}
\end{align}
Here $\mathbf{r}_{\alpha}= \mathbf{x}_{\alpha}+ \bm{b}_\alpha$ is a point on the 
surface $\mathcal S_\alpha$ of the $\alpha$th colloid, where $\mathbf{x}_\alpha$ 
is the location of the center of the $\alpha$th particle and
{$\bm{b}_\alpha$} is the radius vector of the $\alpha$th particle, while $n_k$ is the $k$th component of the outward normal of the  
surface.  The force per unit
area (or the traction) $\mathbf{f}(\mathbf{r}_\alpha)$ on the $\alpha$-th particle is:
${f}_i(\mathbf{r}_\alpha)= \sigma_{ij}n_j$.
 The kernels in the boundary integral representations are
the Green's function $\mathbf{G}$, and the stress tensor $\mathbf{K}$.
They satisfy the Stokes equation \cite{pozrikidis1992}:
\begin{gather}
 {\nabla}_{i} {G}_{ij}\left(\mathbf{r} \right)=0,\\
- {\nabla}_i {P}_j\left(\mathbf{r} \right)+\eta\nabla^{2} {G}_{ij}\left(\mathbf{r} \right)=-\delta\left( \mathbf{r} \right)\delta_{ij},\nonumber\\
{K}_{ijk} ( \mathbf{r} )=-\delta_{ik} {P} ( \mathbf{r})_j+\eta
\big( 
{\nabla_i} {G}_{jk} ( \mathbf{r} )+ {\nabla_k} {G}( \mathbf{r})_{ij}
\big).\nonumber
\end{gather}
{Here ${P}_i$ is the $i$th component of the pressure tensor (Green's function for the pressure) \cite{pozrikidis1992} 
and $\delta \left(\mathbf{r} \right) $ is the Dirac delta function.}
We note that the expression of flow in Eq.\eqref{eq:BIE-PBC} is not absolutely convergent. 
A systematic method to render the flow expression absolutely convergent is well-known \cite{glendinning1982pairwise, brady1988dynamic, ishikawa2008development} by adding extra contributions, which are included in appendix 
\ref{app:convergence}
for completeness.
\begin{figure*}[t!]
\centering\includegraphics[width=0.94\textwidth]{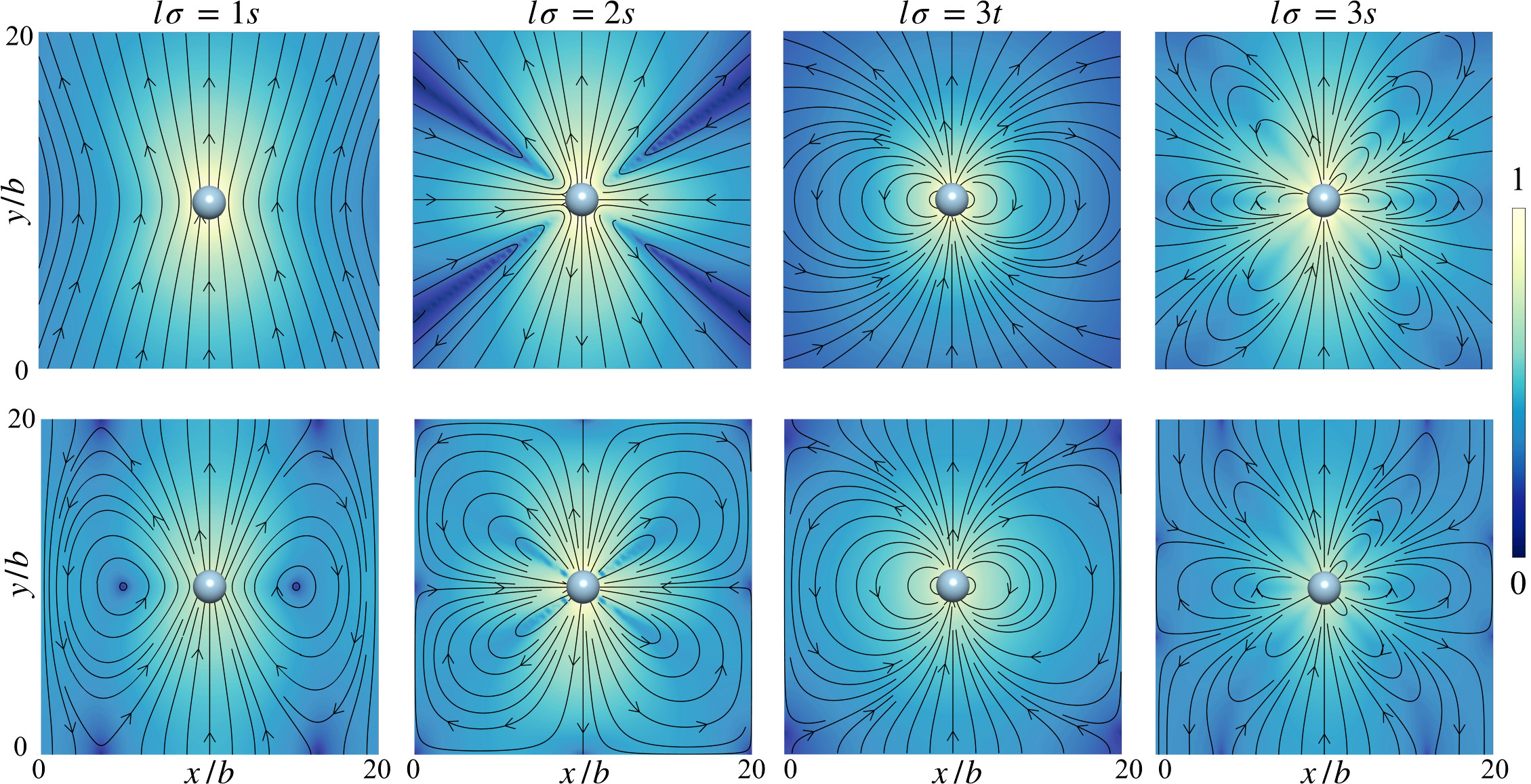}
\caption{{Streamlines of flow overlaid on the pseudocolor plot
of normalized logarithm of the flow speed.}
The panels on top are plots of flow from a given irreducible mode in the unbounded domain, while the panels on the bottom are corresponding flow in the periodic domain using Eq.\eqref{eq:active_Stokes_flow_resolved_in_irreds}.   
The modes $2s,3t,3s$ are parameterized uniaxially as described in Eq.\eqref{eq:uniax} in terms of the orientation (which is chosen along $\hat {y}$),  while $\mathbf F^{e}$ is also chosen along $\hat y$ direction for the $1s$ plot. 
}
\label{fig:flowPeriodicPolar}
\end{figure*}

\subsection{Formal expression of the irreducible components of the flow  in periodic geometry}
We now present the solution of the boundary integrals for the fluid flow following the results in \cite{singh2015many,turk2024fluctuating}, where the results were derived for an unbounded domain and flow bounded by a wall. In periodic domain, we need to do a summation of the flow due to periodic images, as described below. 
The expression for fluid velocity is
obtained from Eq.(\ref{eq:BIE-PBC}) using the method of Galerkin discretization, where we expand the boundary fields in an expansion basis, perform a Taylor expansion of kernels (the Green's function and stress tensor), and use the orthogonality of basis functions to exactly compute the boundary integrals \cite{singh2015many}. 
Our choice of expansion basis is TSH (tensorial spherical harmonics), which are
defined as:
 $\mathbf{Y}^{(l)}(\hat{\boldsymbol{ b }})
   = (-1)^l b^{l+1} \boldsymbol \nabla^{(l)}
  [1/{b}]$  
  \cite{hess2015tensors}.
The expansion of the active slip $\boldsymbol{v}^{\mathcal{A}}$ - defined by Eq.\eqref{eq:slip} - in the basis of the TSH is given as:
\begin{align}
\label{eq:activeVW}
\boldsymbol{v}^{\mathcal{A}}(\boldsymbol{ b }_{\alpha}) 
& =
\sum_{l=1}^{\infty}\tfrac1{(l-1)!(2l-3)!!}\,\mathbf{V}_{\alpha}^{(l)}\cdot\mathbf{Y}^{(l-1)}(\hat{\bm{b}}_{\alpha}).
\end{align}
In the above, and throughout the paper, a dot product denotes  maximal contraction of two tensors. 
The expansion coefficients $\mathbf{V}_{\alpha}^{(l)}$ are irreducible in $l-1$ indices and can be written in terms of irreducible tensors $\mathbf{V}_{\alpha}^{(l\sigma)}$, which are
are symmetric irreducible tensors of rank $l$, $l-1$ and $l-2$ for $\sigma=s,a,$
and $t$ respectively \cite{turk2024fluctuating}.  
The active 
self-propulsion velocity $\mathbf{V}_{\alpha}^{\mathcal{A}}$ 
and self-rotational angular velocity $\bm{\Omega}_{\alpha}^{\mathcal{A}}$ 
 of an isolated active particle due to the slip is
\cite{anderson1991,stone1996}:
\begin{subequations}
\label{eq:VAWA}
\begin{align}
\mathbf{V}_{\alpha}^{\mathcal{A}} & =-\frac{1}{4\pi b^{2}}\int\boldsymbol{v}_{\alpha}^{\mathcal{A}}(\boldsymbol{ b })\,\mathrm{d}\mathcal{S}_{\alpha},
\\
\boldsymbol{\boldsymbol{\Omega}}_{\alpha}^{\mathcal{A}} & =-\frac{3}{8\pi b^{4}}\int\boldsymbol{ b }_{\alpha}\times\boldsymbol{v}_{\alpha}^{\mathcal{A}}(\boldsymbol{ b })\,\mathrm{d}\mathcal{S}_{\alpha}.
\end{align}
\end{subequations}

\begin{figure}[t!]
\centering\includegraphics[width=0.45\textwidth]{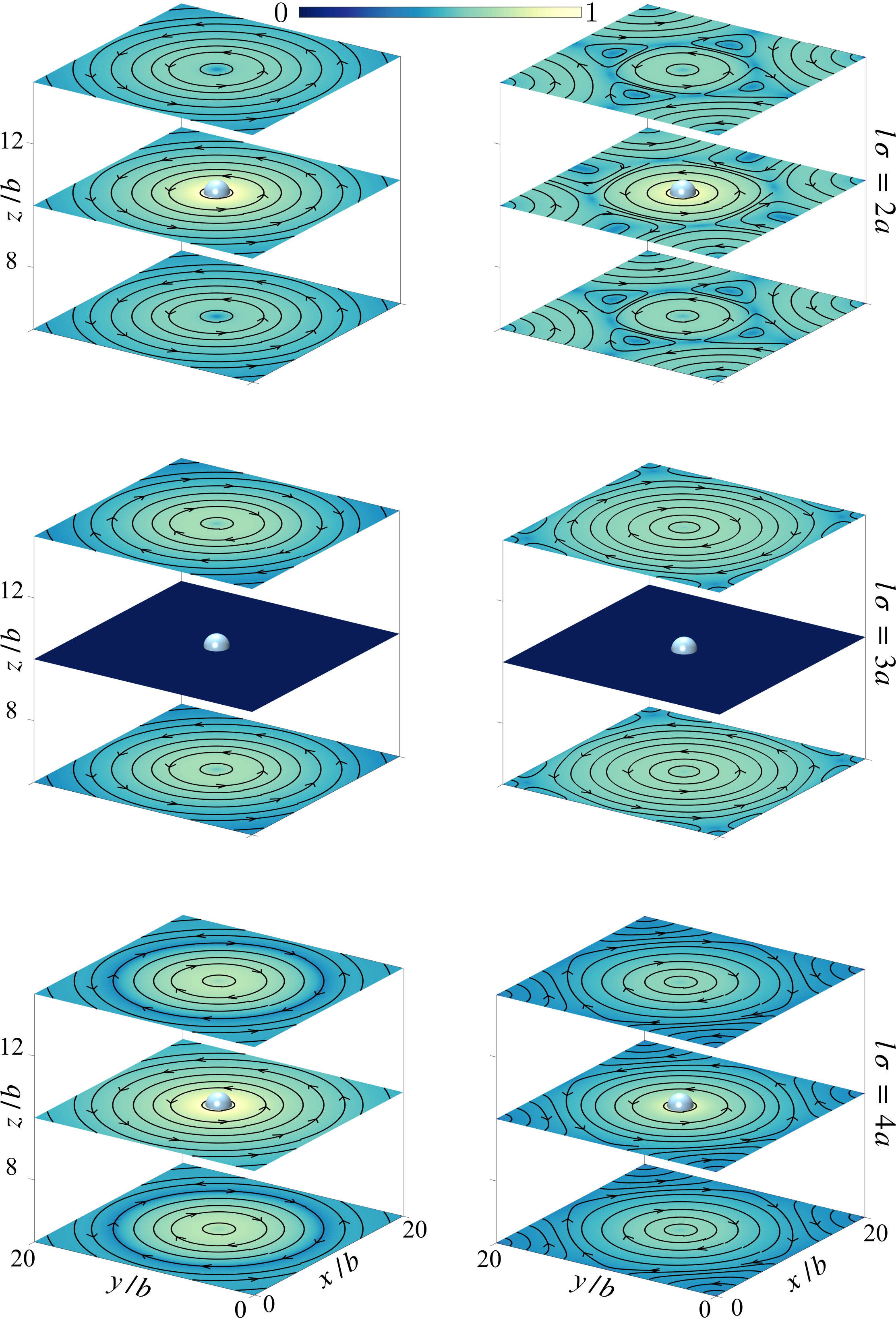}
\caption{{Streamlines of swirling fluid flow - see Eq.\eqref{eq:active_Stokes_flow_resolved_in_irreds} -  overlaid on the
pseudocolor plot of normalized logarithm of the flow speed. }
Top row is due to a rotlet ($l\sigma=2a$) with a net body torque along $\hat{\boldsymbol z}$. The middle row is due to the mode $l\sigma=3a$. The last row is due to $l\sigma=4a$. Panels on LEFT are flow in an unbounded domain , while the panels on the RIGHT are corresponding flow due to irreducible modes in the periodic domain. 
The modes $3a,4a$ are parameterized uniaxially as described in Eq.\eqref{eq:uniax} in terms of the orientation of the particle, which is chosen to be along $\hat z$ direction. For the $2a$ mode, the body torque is along $\hat z$ direction.
\label{fig:flowPeriodicChiral}}
\end{figure}

The expression of the fluid flow due to active colloidal particles in periodic domain can be written as the following summation over irreducible parts $\boldsymbol v^{l\sigma}$:
\begin{align} \label{eq:active_Stokes_flow_resolved_in_irreds}
\boldsymbol{v} (\mathbf{x})
  &= \sum_n\sum_{\alpha=1}^N
   \sum_{l\sigma=1s}^\infty
 \boldsymbol v^{l\sigma}  (\mathbf{x},\mathbf{x}_{\alpha,n}).
\end{align}
Here, $\sigma\in\{s,a,t\}$ for each $l$ (which goes from 1 to $\infty$), while sum over the index $n$ implies summation over the periodic images.
%
We now write formal expressions of irreducible part to clearly separate the active and the passive terms. 
The passive part of the fluid flow is due to $l\sigma=1s$ (a net body force $\mathbf F^e$) and 
$l\sigma=2a$ (a net body torque $\mathbf T^e$). Their formal expressions are:
\begin{align}
 \boldsymbol v^{1s} =   \mathcal{F}^{0}\mathbf{G} \cdot \mathbf{F}_\alpha^e,\qquad
\boldsymbol v^{2a} =   \tfrac12  \left( 
\boldsymbol\nabla_{\alpha} \times \mathbf{G} 
\right) 
\cdot \mathbf{T}_\alpha^e.
    \label{eq:passiveFLows}
\end{align}
Here, the derivative $\boldsymbol \nabla_\alpha$ implies that derivative has been taken with respect to the vector $\mathbf{x}_\alpha$, which is the position vector of the $\alpha$-th particle. 
We have also defined the finite-size operator, which encodes the finite radius $b$ of the particle: $\mathcal{F}^{l}=1+\frac{b^2}{4l+6}\nabla^2.$
The mode $1s$ is sometimes referred to as Stokeslet and mode $2a$ is referred to as rotlet \cite{grahamMicrohydrodynamics2018}. The flow due to $1s$ mode is  plotted in Figure (\ref{fig:flowPeriodicPolar}) for both unbounded and periodic geometry of Stokes flow. The flow due to $2a$ mode is  plotted in Figure (\ref{fig:flowPeriodicChiral}) for both unbounded and periodic geometry of Stokes flow.

The active velocity fields are from the arbitrary slip mode $\mathbf{V}_\alpha^{(l\sigma)}$ are given as
\cite{singh2015many,singh2019competing,turk2024fluctuating}:
\begin{align}
    \boldsymbol v^{ls} &= c^s_l\,\, \,\mathcal{F}^{l-1}\,\,
    \big(
   \boldsymbol{\nabla}_{\alpha}^{(l-1)}\,\mathbf{G}\,
    \big)\cdot \mathbf{V}_\alpha^{(ls)},\nonumber
    \\
    \boldsymbol v^{la} &=c^a_l\left[\boldsymbol{\nabla}_{\alpha}^{(l-2)}
    \left(\boldsymbol{\nabla}_{\alpha}\times\mathbf{G}\right)
     \right]\cdot \mathbf{V}_\alpha^{(la)},\nonumber
    \\
    \boldsymbol v^{lt} &= c^t_l
    \left[\,\boldsymbol{\nabla}_{\alpha}^{(l-3)}\,\,\left({\nabla^2}\,\mathbf{G}\right)
     \right]\cdot \mathbf{V}_\alpha^{(lt)}.
    \label{eq:activeFLows}
\end{align}
Note that $ \boldsymbol v^{ls} $ is defined only for $l\geq2$, while 
$ \boldsymbol v^{la}$ and $ \boldsymbol v^{lt}$ are defined only for $l\geq3$. 
In the above, we have used a notation where $\boldsymbol{\nabla}_{\alpha}^{(l)}$ is a tensor of rank $l$. Explicitly, in the index notation, it can be written as: 
$[{\nabla}_{\alpha}^{(l)}]_{i_1 i_2\dots i_l}=[{\nabla}_{\alpha}]_{i_1}
\,[{\nabla}_{\alpha}]_{i_2}\dots [{\nabla}_{\alpha}]_{i_l}.$ 
The value of the constant coefficients $c_{l}^{\sigma}$ can be obtained from solving the linear system which results from the boundary integral equation \cite{turk2022stokes,turk2024fluctuating}. For the leading modes, these constants are: $c^s_2=10\pi\eta b^2/3$, while
$c^t_3=-2\pi\eta b^3/5$. 
Thus, the expression of the flow can be obtained once a Green's function is specified. The explicit forms of the above irreducible components of the fluid flow is available in appendix \ref{app:EwaldOseen} and \ref{app:IrredFlow}.  
 
In the remainder of the paper, we parameterize the irreducible modes of the slip in terms of TSH. 
This is a natural choice as irreducible modes of the slip are themselves symmetric traceless tensors. The expression for the mode of the slip on the $\alpha$-th particle is written in terms of its orientation $\mathbf{e}_\alpha$ as (here $V_{0}^{(l\sigma)}$ are constants):
    \begin{gather}
     \label{eq:uniax}
      \mathbf{V}_{\alpha}^{(ls)}=V_{0}^{(ls)}\,\mathbf{Y}^{(l)}(\mathbf{e}_{\alpha}),\\ 
    \mathbf{V}_{\alpha}^{(la)}=V_{0}^{(la)}\,\mathbf{Y}^{(l-1)}(\mathbf{e}_{\alpha}),\quad 
    \mathbf{V}_{\alpha}^{(lt)}=V_{0}^{(lt)}\,\mathbf{Y}^{(l-2)}(\mathbf{e}_{\alpha}).\nonumber
    \end{gather}
We note that the above is a uniaxial parametrization of the slip, while in general the irreducible modes of the slip can be of any general form as long as they are symmetric and traceless. Thus, the above choice is suited to active particles which have uniaxial symmetry. For particles lacking this symmetry a more appropriate parametrization should be chosen.  

From the Eq.\eqref{eq:activeFLows}, it is clear that the flow due to the $l\sigma$ mode decays as $r^{-l}$, where $r$ is the distance from the source (since $\mathbf G$ decays as $r^{-1}$, see appendix \ref{app:EwaldOseen}). 
We plot the flow around colloids due to leading modes of activity
by parametrizing the slip modes uniaxially in Figures 
(\ref{fig:flowPeriodicPolar}) and (\ref{fig:flowPeriodicChiral}). 
 Fluid
flow due to $l\sigma=1s,2s$ and $3t$ have been plotted in Figure (\ref{fig:flowPeriodicPolar}). Swirling flows due to modes $l
\sigma=2a$, $3a$ and $4a$ have been plotted in Figure (\ref{fig:flowPeriodicChiral}).
The flow has been plotted in both unbounded (in terms of Oseen tensor and its derivatives) and periodic (using Ewald summation of the Oseen tensor and its derivatives) geometry of Stokes flow for comparison.

The nature of the flow can then be determined from the parity of the
mode. The flow due to the mode $1s$  - a net body force - 
decays as $r^{-1}$.
The mode $2a$, which is due to a net torque on the particle, decays as $r^{-2}$.
The mode $2s$ - which is also called stresslet \cite{grahamMicrohydrodynamics2018} - corresponds to a symmetric irreducible dipole and decays as $r^{-2}$.
The stresslet mode is apolar in nature, and thus, can not lead to self-propulsion of a spherical particle. A polar nature of fluid flow can only arise at $l=3$, where the flow decays as $r^{-3}$. Similarly, the lowest mode which is capable of producing an active spinning of an isolated active particle is $l\sigma=4a$, which decays as $r^{-4}$
\cite{ghose2014irreducible,pak2014generalized,pedley2016squirmers}.
Thus, it follows that the leading flow behavior from a pure
self-propulsion is $r^{-3}$, while that from a pure self-rotation is $r^{-4}$.

\section{Rigid body motion of hydrodynamically interacting active particles} \label{sec:RBM}
Having obtained the irreducible components of the fluid velocity, we now turn to compute the expression for rigid body motion of active particles for dynamical simulations. 
This is indeed the most important quantity as these can be used to run numerical simulations and compute rheological quantities. The self-propulsion and rotation of an isolated active particle
due to surface slip in an infinite system is given in Eq.\eqref{eq:VAWA}. We can compute hydrodynamic interactions using the boundary integral method. The rigid body motion of active colloids in terms of the known coefficients of the slip,
body forces and torques has been obtained for unbounded domain and flow bounded by a plane surface \cite{singh2015many,turk2024fluctuating}. 
The result for the periodic domain can be written as:
\begin{subequations}
\label{eq:rigidBody}
    \begin{align}
\mathbf{U}_\alpha & =\mu_{t}\,\mathbf F^e_\alpha 
+ \pi_{t}\, \mathbf{V}_\alpha^{\mathcal{A}} + 
\sum'_n
{\sum_{\alpha=1}^{N}}\,
{\sum_{\beta=1}^{N}}
 \sum_{l\sigma=1s}^\infty
\boldsymbol U_{\alpha\beta}^{l\sigma},\\
\boldsymbol{\Omega}_{\alpha} & = \mu_{r}\, \mathbf T^e_\alpha 
+ \pi_{r}\,\boldsymbol \Omega_\alpha^{\mathcal{A}}
+ 
\sum'_n
{\sum_{\alpha=1}^{N}}\,
{\sum_{\beta=1}^{N}}
\sum_{l\sigma=1s}^\infty
\boldsymbol W_{\alpha\beta}^{l\sigma}.
\end{align}
\end{subequations}
Here, $\sigma\in\{s,a,t\}$ for each $l$, the prime over the summation index $n$ for periodic images implies that self-interaction ($\alpha=\beta$) is not computed from the original box $n=1$, but it is computed from other periodic images ($n\neq 1$). 
In Eq.\eqref{eq:rigidBody}, we have defined the following scalars:  $\mu_t =   \frac{1}{6\pi\eta b}-\frac{\xi}{\pi^{3/2}\eta }+\frac{20\xi^3b^2}{9\eta\pi^{3/2}},$ 
$\pi_t=1-\frac{40\xi^3 b^3}{3\sqrt{\pi}}$,  
$ \mu_r =   \frac{1}{8\pi\eta b^3}
 -\frac{5\xi^3}{3\eta\pi^{3/2} }$, and $\pi_r=1$.
The contributions proportional to $\xi$ are due to self-contributions which stem from the Ewald summation procedure  \cite{beenakker1986ewald}. They vanish in the limit of $L\rightarrow \infty$.

We now provide formal expressions of $\boldsymbol U^{l\sigma}_{\alpha\beta}$ 
and $\boldsymbol W^{l\sigma}_{\alpha\beta}$ which are given in Eq.\eqref{eq:rigidBody}.
We first given expressions for $1s$ and $2a$ modes, which correspond to
body forces and torques on the particles. These are given as \cite{ladd1988,mazur1982,brady1988dynamic} :
\begin{subequations}
    \label{eq:mobMat}   
\begin{align}
   \boldsymbol U_{\alpha\beta}^{1s}&=  \mathcal{F}^{0}\mathcal{F}^{0}\mathbf{G}\cdot \mathbf F_\beta^e,\\
    \boldsymbol U_{\alpha\beta}^{2a}
    &= 
     \dfrac12 
     {
   \left( \boldsymbol\nabla_\beta \times \mathbf{G}\right)
    \cdot \mathbf T_\beta^e},
    \\
     \boldsymbol W_{\alpha\beta}^{1s}&=
     \dfrac12 { 
    \boldsymbol\nabla_{\alpha}\times 
     \left(\mathbf{G}
     \cdot \mathbf F_\beta^e\right)},\\
    \boldsymbol W_{\alpha\beta}^{2a}&=
     -\dfrac1{4} {\nabla^2 \mathbf{G} \cdot \mathbf T_\beta^e}.
\end{align}
\end{subequations}
In the above, $\boldsymbol\nabla_{\alpha}$ implies that the derivative has been take with respect to $\mathbf{x}_{\alpha}$ and the sum over the particle index $\beta$ is implied. 
Note that the Green's function depends on $\mathbf{x}_{\alpha}-\mathbf{x}_{\beta}$. Thus, 
$\boldsymbol\nabla_{\alpha}=-\boldsymbol\nabla_{\beta}$. But the Laplacian does not depend on the particle index in periodic domain which has translation invariance: $\nabla^2_{\alpha}=\nabla^2_{\beta}=\nabla^2$.
Ewald summed expressions of the above are given in appendix \ref{app:RBM}. 
The contributions to hydrodynamic interactions due to the active slip can also be written in terms of the derivatives of the Green's function. The results are
\cite{singh2015many,turk2024fluctuating}:
\begin{subequations}
\begin{align}
    \boldsymbol U_{\alpha\beta}^{ls} &=c^s_l 
 \mathcal{F}^{0}\mathcal{F}^{l}
    \big[\boldsymbol{\nabla}_{\beta}^{(l-1)}\mathbf{G}\big]
    \cdot \mathbf{V}^{(2s)}_{\beta},
    \\
   \boldsymbol U ^{la}_{\alpha\beta}&
   = c^a_l
   \left(\boldsymbol{\nabla}_{\beta}^{(l-2)}
    [\boldsymbol{\nabla}_{\beta}\times\mathbf{G} \,]
    \right)\cdot \mathbf{V}^{(la)}_{\beta} ,
    \\ 
    \boldsymbol U^{lt}_{\alpha\beta}&=
    c^t_l
    \left(\boldsymbol{\nabla}_{\beta}^{(l-3)}[{\nabla^2_\beta}\, \mathbf{G}]
    \right)\cdot \mathbf{V}^{(lt)}_{\beta},
    \\
\boldsymbol W^{ls}_{\alpha\beta} &=
\frac{c^s_l}2
\boldsymbol\nabla_\alpha \times 
\left(
\big[\,\boldsymbol{\nabla}_{\beta}^{(l-1)}\mathbf{G}
\big]\cdot \mathbf{V}^{(ls)}_{\beta}\right),
  \\
     \boldsymbol W^{la}_{\alpha\beta}&= 
     \frac{c^a_l}2
     \boldsymbol\nabla_\alpha
     \times
        \left(   
        \big[ 
        \boldsymbol{\nabla}_{\beta}^{(l-2)}
    \boldsymbol{\nabla}_{\beta}\times\mathbf{G} 
    \big]
    \cdot \mathbf{V}^{(la)}_{\beta}
    \right).
 \label{eq:PropTensors}
   \end{align}
\end{subequations}
The contributions due to mode $ls$ are defined for $l\geq 2$, while the terms due to mode $la$ and $lt$  are defined for $l\geq 3$.
We note that $\boldsymbol W_{\alpha\beta}^{lt}=0$ as the Stokes equation ensures that $\boldsymbol\nabla\times \nabla^2 \mathbf G =0$.
Or, the fact that the curl of the Laplacian of a valid Stokes flow vanishes.
Ewald summed expression of the $\boldsymbol U^{l\sigma}_{\alpha\beta}$ 
and $\boldsymbol W^{l\sigma}_{\alpha\beta}$ is given explicitly in appendix \ref{app:RBM}. These explicit expression can be used to run numerical simulations of active particles in periodic geometries. These expression for rigid body motion (appendix \ref{app:RBM}) and fluid flow (appendix \ref{app:IrredFlow}) for terms up to $l\sigma=4a$ are central results of this paper.
To the best of knowledge, explicit forms of these contributions to flow and rigid body motion have not been reported elsewhere for terms beyond $l=2$.

\begin{figure*}[t!]
\centering\includegraphics[width=0.94\textwidth]{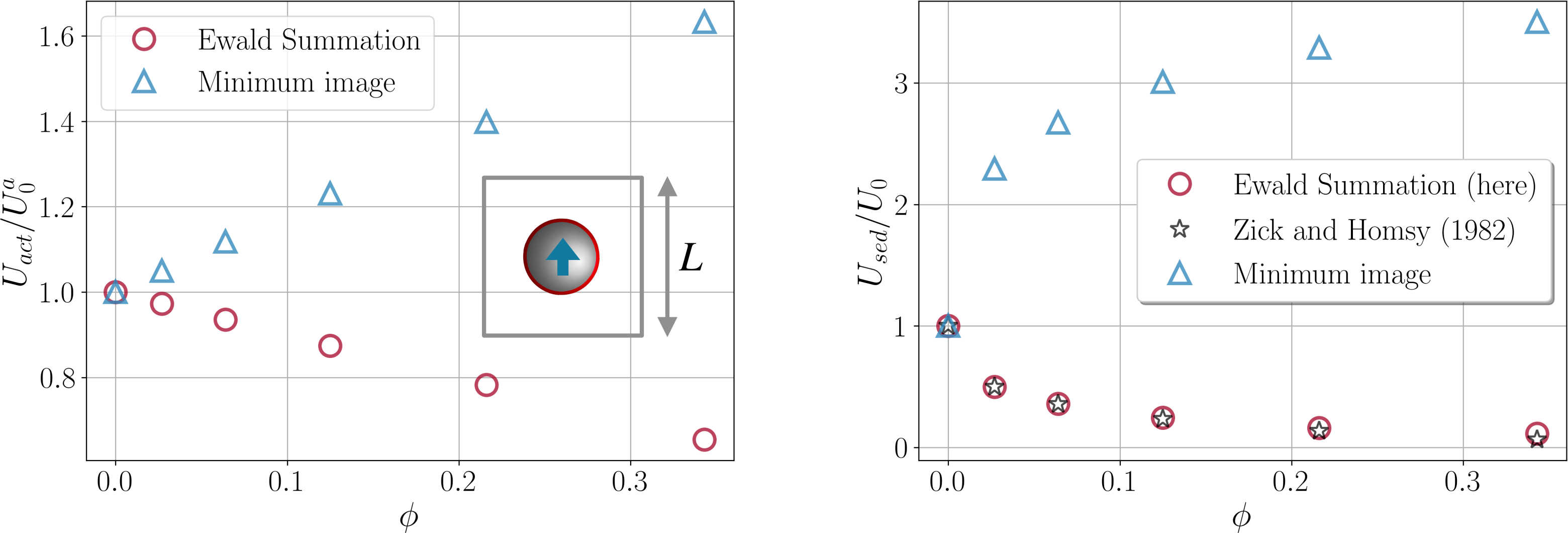}
\caption{LEFT: Speed of an active sphere in a periodic three-dimensional simulation box, a cube of size $L$, computed using minimum images and Ewald summation (see schematic in the inset). The speed is plotted as a function of volume fraction $\phi$. Here $U^a_0$ is the self-propulsion speed of an isolated active particle  (the limit $L\rightarrow \infty$). Note that the flow due to a self-propelling colloid is due to a $3t$. Thus, it is a higher order mode. RIGHT: Hindrance in the speed of a sedimenting sphere ($U_{sed}$) due a 
net body force (gravity) on the particle as a function of volume fraction $\phi$. Here $U_0$ is the speed of an isolated sphere (the limit $L\rightarrow \infty$). Ewald summation (here) is shown, which is compared with result of minimum image convention. Existing result from literature is also shown (Zick and Homsy (1982) \cite{zick1982stokes}). 
\label{fig:EwaldMin}}
\end{figure*}

Having obtained expression for the rigid body motion, we can now run dynamical simulations of active particles in periodic geometry of Stokes flow. The position $\mathbf{x}_\alpha$ and orientation $\mathbf{e}_\alpha$ of $\alpha$-th particle are updated using Eq.\eqref{eq:rigidBody} from the following kinematic equations:
\begin{align}
\dot{\mathbf{x}}_\alpha =   \mathbf U_\alpha,\qquad \dot {\mathbf{e}}_\alpha = \boldsymbol \Omega_\alpha\times \mathbf{e}_\alpha.
\end{align}
Thus, we have shown above the the problem of studying collective behavior of active particles in periodic geometry has been reduced to choice of the slip
and to obtain the derivatives of the periodic Green's function. In the next section, we present some representative applications of our method.

\section{Comparison of full Ewald summation to minimum images}\label{sec:app}
{
The Ewald summation is an efficient computational technique which accelerates the summation of periodic images. In this section, we compare the Ewald summed solutions to the one with minimum image convention (where only nearest periodic images are considered) for one active and one passive system. 
 In section \ref{app:activeLattice}, we apply our theory to compute the velocity of a simple cubic lattice of active particles. This is a higher order mode ($3t$). Still, we find that the minimum image does not predict the hindered motion of the lattice with increasing volume fraction. In section \ref{app:sedimentingLattice}, we show the classic example of sedimenting cubic lattice ($1s$ mode in our notation) and compare our result using Ewald summation with existing literature. For both these representative examples, 
 the simple minimum image gives both qualitative and quantitative wrong results. 
 }


{\subsection{A simple cubic lattice of active particles}\label{app:activeLattice}
In this section, we study the speed of cubic lattice of active particles (the activity is due to the $3t$ mode of the active slip, which is responsible for active self-propulsion). The $3t$ is also called source dipole \cite{lauga2020fluid}. The system of a simple cubic lattice of active particles can be modelled as a sphere in a box of length $L$ along with periodic images. In an infinite system ($L\rightarrow\infty$), the self-propulsion speed of this particle is chosen to be $U_0^a$. Thus, its active velocity is $\mathbf V^{\alpha}=U_0^a\,\mathbf e$, where $\mathbf e$ is its orientation.  
We now seek to compute the speed of a simple cubic lattice of active particles using minimum-image and full Ewald summation. As described above, the flow due to force-free active sphere which self-propels is possible due to the $3t$ mode. Using Eq.\eqref{eq:rigidBody}, 
the effective active velocity $\mathbf U_{act}$ is:
\begin{align}
    \mathbf U_{act} = \pi_t\,\mathbf V^{\mathcal{A}} + \sum_n'  \bm U^{(3t)}
   \end{align}In the above, the prime on the summation indicates that self-contribution from the main box is excluded and only contributions from periodic images is considered. Explicit form of $\bm U^{(3t)}$ is available in appendix \ref{app:3tRBM}. In addition $\mathbf{V}_\alpha^{(3t)}$ is proportional to proportional to the self-propulsion velocity. 
   Thus, we have \cite{turk2022stokes,turk2024fluctuating}:
    $\mathbf{V}^{(3t)} = 5\mathbf{V}^{(A)}$.
Using the above information, we can use the Ewald summation of the $3t$ and compare it with the minimum image convention. See Fig.\ref{fig:EwaldMin}. It can be seen that the minimum images and full Ewald summation only matches in the limit of $L\rightarrow\infty$. Thus, it is essential to do Ewald summation for polar $3t$ mode, as it decay as $1/r^3$, which is long-ranged in three-dimensions. }

{\subsection{A simple cubic lattice of sedimenting particles}\label{app:sedimentingLattice}
In this section, we use the formalism developed in the paper to compute the sedimentation speed of a simple cubic array of spheres as a function of volume fraction $\phi=N\mathcal{ V}_{\text{sph}}/\mathcal V$. Here $\mathcal{ V}_{\text{sph}}=4\pi b^3/3$ is the volume of a single sphere of radius $b$, while $N$ is the number of spheres in a box of volume $\mathcal{ V}=L^3$. 
Our results are compared with results present in literature, with the paper of Zick and Homsy \cite{zick1982stokes}. In our simulation, we have a single sphere in a box of size $L$ and we vary $L$ to change volume fraction $\phi$. Ewald summation is used to compute interaction with periodic images.  
In addition, we show that the minimum image (where the spheres in the main box only interacts with nearest images) predicts the velocity of the lattice should increase with volume fraction. On the other hand, the correct picture shows hindered setting. 
Indeed, a system of two spheres falls faster than one sphere, and thus, a minimum image shows the speed to increase.}

{In the presence of periodic images, using Eq.\eqref{eq:rigidBody}, the velocity of the sedimenting particle is:
\begin{align}
    \mathbf U_{sed} = \mu_t\,\mathbf F_0 + \sum_n'  \bm U^{(1s)}.
\end{align}
Here $\mathbf F_0$ is the net gravitation body force on the particle, while the summation is over periodic images of the system. In the above, the prime on the summation indicates that self-contribution from the main box is excluded and only contributions from periodic images is considered. Explicit form of $\bm U^{(1s)}$ is available in appendix \ref{app:1sRBM}.
Thus, we have an infinite lattice of spheres, which is known to show hindered sedimentation \cite{zick1982stokes, zhangMenonPRL2025}. To model this, we need Ewald summation, which shows that the velocity decreases with volume fraction $\phi$. See Fig.(\ref{fig:EwaldMin}) for the results. This phenomena of hindered sedimentation is not captured by minimum image convention. }

{To summarise, in this section we have demonstrated that even for higher order terms, such as the $3t$ mode, a full Ewald summation is crucial in three-dimension. This is because the flow at $l=3$ decays as $1/r^3$. Thus, all terms up to $l=3$ are truly long-ranged in three-dimensions, and an Ewald summation is required for these modes. Consequently, the minimum image convention is not expected to converge.  A comprehensive study which compares the minimum image to the full Ewald summation for all modes of the active slip when quantities - such as volume fraction and strength of the activity - are varied, suggests an exciting direction for future work. }
    
\section{Summary and discussion\label{sec:Conclusion}}
In this work, we have presented a method to compute the many-body
hydrodynamic interaction of $N$ spherical colloids with active boundary
layers in periodic geometry of Stokes flow. The solution is obtained from gradients of the Ewald summation of the Oseen tensor. Explicit expression has been provided for terms up to $l=4$ in expansion basis. Thus, the method and explicit expression of the paper provides an
efficient tool to study dynamics of active colloids in an irreducible
basis and for a rheological study of active suspensions.

The method presented in this paper uses a direct method for the computation
of hydrodynamic interactions. This requires an $O(N^{2})$ computational
effort. The cost can be reduced to $O(N\log N)$ by using
methods such as particle-mesh Ewald \cite{darden1993particle}.
In this work, we have only considered hydrodynamic interactions and ignored the role of chemical interactions between the colloids and their correlated Brownian motion \cite{sprinkle2017large}. It is possible to extend our method to study 
the competing roles of chemical and hydrodynamic interactions  \cite{singh2019competing}, and also the Brownian motion of active colloids \cite{turk2024fluctuating}. 
Although, our theory is for spherical particles, it can be used to simulate extended structures such active polymers \cite{laskar2015brownian,winklerPhysicsActivePolymers2020} and active sheets \cite{manna2022harnessing} in periodic geometry of Stokes flow. 
{In this paper, we have assumed periodicity in all three directions for the Ewald summation procedure. Another exciting direction is to extend our work for periodicity in two directions \cite{bleibel2012ewald,pozrikidis1996computation}}.
All these 
suggest exciting direction for future research. 

\section*{Acknowledgments}

We would like to thank R. Adhikari, ME Cates, 
and G. Turk for seminal discussions. We also thank two anonymous reviewers
for their comments, which has led to
an improvement in the presentation of our results.
R.S. acknowledges
support from the Indian Institute of Technology, Madras, India and their seed and initiation grants as well
as a Start-up Research Grant, SERB, India (SERB file number: SRG/2022/000682).

\appendix
\section{Ewald summation of Oseen tensor}\label{app:EwaldOseen}
The Green's function $\mathbf{G}(\mathbf{r})$ of Stokes flow in the unbounded domain is \cite{pozrikidis1992}:
\begin{align}
8\pi \eta\,{G}_{ij}(\mathbf{r})= \frac{\delta_{ij}}{r} + \frac{r_i r_j}{r^{3}}   
=\left(\delta_{ij}\nabla^{2}-{\nabla_i}{\nabla_j}\right)r.
\label{eq:Oseen}
\end{align}
The above is often refereed to as the Oseen tensor in the literature. 
In periodic suspensions, one needs to perform lattice sum
of the Oseen tensor from the periodic images.
The Oseen tensor decays as $r^{-1}$, and thus, the lattice sums converge
slowly and a direct summation is not suitable for numerical simulations. To this end, we compute the Ewald summation of the Oseen tensor using \cite{beenakker1986ewald,pozrikidis1992}.
 Following
\cite{beenakker1986ewald}, we have separated the above expression
of Oseen tensor in parts which decay rapidly in real and Fourier space respectively. They are:
%
\begin{alignat}{1}
\mathbf{G}(\mathbf{r}) & =\frac{1}{8\pi\eta}
\left(\mathbf{I}\,\nabla^{2}-\boldsymbol{\nabla}\boldsymbol{\nabla}\right)\big[
r\,\mathrm{\mathrm{erfc} }(\xi r)
+r\,\mathrm{erf }(\xi r)\big]
\label{eq:Gerf}
\end{alignat}
In the above, $\xi$ is a positive constant with dimensions of inverse length. In this paper, the simulations are performed in a cube of side $L$ and we set the optimal value of $\xi = \sqrt{\pi}/L$
such that the sum converges rapidly \cite{nijboer1957calculation,beenakker1986ewald}. Here, we have defined:
\begin{align}
\mathrm{erf}(q) = \frac{2}{\sqrt{\pi}} \int_0^q e^{-t^2}\mathrm{d}t,\quad 
\mathrm{erf}(q) + \mathrm{erfc}(q)=1.
\end{align}

In Eq.\eqref{eq:Gerf} we have defined the Green's function of Stokes flow and written them in terms of parts decaying rapidly in real and Fourier spaces. They are:
\begin{align}
  8\pi\eta\,  \mathbf{G}^{\mathbb R}
(\mathbf{r}) &=  \left(\mathbf{I}\,\nabla^{2}-\boldsymbol{\nabla}\boldsymbol{\nabla}\right) r\, \mathrm{\mathrm{erfc} }(\xi r) 
,\\
8\pi\eta\, \mathbf{G}^{\mathbb F} (\mathbf{r}) &=\left(\mathbf{I}\,\nabla^{2} -\boldsymbol{\nabla}\boldsymbol{\nabla}\right) r\, \mathrm{\mathrm{erf} }(\xi r). 
\label{eq:GerfExp}
\end{align}
The function $\mathbf{G}^{\mathbb R}(\mathbf{r})$ decays rapidly in real space as a function of distance $r$, while the function $\mathbf{G}^{\mathbb F}(\mathbf{r})$ decays slowly in the real space (and thus rapidly in the Fourier space). We utilize this information for summation of the Green's function and do the Ewald summation in both real and Fourier space \cite{beenakker1986ewald}.
The rapidly decaying function in the real space will converge
very quickly while the term which is slowly varying in real space
can be rendered convergent in the Fourier space with very less number
of Fourier modes.  

We define the Fourier transform of a function $\varphi(\mathbf{r})$
as:  
\begin{align}
    {\hat \varphi}(\mathbf{k})&=\int\varphi(\mathbf{r})e^{-i\mathbf{k}\cdot\mathbf{r}}d\mathbf{r},\quad\,
\varphi(\mathbf{r})=\frac{1}{(2\pi)^{3}}\int \hat {\varphi}(\mathbf{k})e^{i\mathbf{k}\cdot\mathbf{r}} d\mathbf{k}.\nonumber
\end{align}
    The
transformation to the Fourier space is possible due to the Poisson
summation formula, which relates sum over lattice vectors $\mathbf{r}_{n}$
to the reciprocal lattice vectors $\mathbf{k}_{n}$  for a function $\varphi(\mathbf{r}_{n})$ as:
$\sum_{n}\varphi(\mathbf{r}_{n})=\frac{1}{\mathcal V}\sum_{n}\hat \varphi (\mathbf{k}_{n})$
and by calculating the Fourier transform of the Oseen tensor, as
we now describe. The Fourier transform of the Oseen tensor, Eq. (\ref{eq:Oseen}),
is:
\begin{align}
    \hat{G}_{ij}^{\mathrm{ }}(\mathbf{k})= 
    \frac{1}{\eta k^2}
    \left(
    \delta_{ij} -  \hat{k}_{i}\hat{k}_{j}
    \right).    
\end{align}
Here $\hat k_i = k_i/k$, while $k=|\mathbf{k}|$. 
Using this expression in Eq.(\ref{eq:GerfExp}), the expressions for $\mathbf{G}^{\mathbb R}$
and $\hat{\mathbf{G}}^{\mathbb F}$ becomes:
\begin{subequations}
\begin{align}
\label{eq:GrGf}
8\pi\eta \,\mathbf{G}^{\mathbb R}(\mathbf{r})
&=A(\xi r)\frac{\mathbf{I}}{r}+B(\xi r)\frac{\hat{\mathbf{r}}\hat{\mathbf{r}}}{r},\\
8\pi\eta \,\hat{\mathbf{G}}^{\mathbb F}(\mathbf{k}) 
&=C( k/\xi)\,\cos(\mathbf{k}\cdot\mathbf{r})\,\left(\frac{\mathbf{I}-\hat{\mathbf{k}}\hat{\mathbf{k}}}{k^{2}}\right).
\end{align}
\end{subequations}
Here, we have made the following definitions: 
\begin{subequations}
\label{eq:ABC}
\begin{align} 
    A(q)&= \mathrm{\mathrm{erfc}}(q)+\left({2q^{3}-3q}\right)\frac{2e^{-q^2}}{\sqrt{\pi}} ,\\
B(q)&= 
\mathrm{\mathrm{erfc}}(q)+\left(q-2q^{3}\right)\frac{2e^{-q^2}}{\sqrt{\pi}}  ,\\ 
C(q)&=\frac{8\pi}{ \mathcal{V}}\left[1+\frac{q^2}{4} +\frac{q^4}{8} \right]
\exp\left(
-\tfrac14 q^2
\right).
\end{align}
\end{subequations}
\section{Irreducible components of flow}
\label{app:IrredFlow}
In the following, we present the Ewald summation of the irreducible components of flow around an active particle obtained from the boundary integral representation of the Stokes flow. 
The formal expressions of the irreducible components of the flow is given in Eq.\eqref{eq:passiveFLows}-\eqref{eq:activeFLows} of the main text. In this section, we give explicit forms of flow due to a single particle from modes $1s$, $2s$, $2a$, $3s$, $3a$, $3t$, $4a$. For clarity, we write real and Fourier part of the flow as  $[{v^{l\sigma}}]^{\mathbb R}_i$ and $[{v^{l\sigma}}]^{\mathbb F}_i$ such that:
    \begin{align}
 {v_i^{l\sigma}}(\mathbf{x},\mathbf{x}_\alpha)=
[{v_\alpha^{l\sigma}}]^{\mathbb R}_i(\mathbf{r} ) +  [{v_\alpha^{l\sigma}}]^{\mathbb F}_i( \mathbf{k}_n)  
    \end{align}
    In the above, the distance from the source point is defined as: $r=|\mathbf{r}|$, while $\mathbf{r} = \mathbf{x}-\mathbf{x}_{\alpha,n}$, while the sum $n$ is over all periodic images and wave vectors.
\subsection{Flow from the $1s$ mode}
\label{sec:flow1s}
In index notation, the flow due to the $1s$ mode is 
$$v_i^{1s}(\mathbf{x},\mathbf{x}_\alpha)=\mathcal{F}^0\,G_{ij}F^e_{\alpha,j}.$$ 
The velocity components are:
\begin{align*}
   [{v_\alpha^{1s}}]^{\mathbb R}_i&= 
   \frac{1}{ 8\pi\eta}
     \left(A(\xi r) \frac{\delta_{ij}}{r} + B(\xi r)\frac{r_{i} r_j}{r^{3}}\right) F_{\alpha,j}^e
     \nonumber\\
     &+    \frac{1}{ 8\pi\eta}
\frac{b^2}{6} 
          \left(\tilde A(\xi r) \frac{\delta_{ij}}{r^3} + \tilde B(\xi r)\frac{r_{i} r_j}{r^{5}}\right) F_{\alpha,j}^e
          \\
[{v_\alpha^{1s}}]^{\mathbb F}_i &=   
\left(  1 - \frac{b^2k^2}{6}  \right) \hat G^{\mathbb F}_{ij}(\mathbf{k})\,F_{\alpha,j}^e
\
\end{align*}
In Eq.\eqref{eq:ABC}, $A(q)$ and $B(q)$ was defined. Here we give
\begin{align}
    \tilde A(q) &=  2\mathrm{\mathrm{erfc}}(q)+
\frac{2e^{-q^2}}{\sqrt{\pi}}
\left[2q + 28 q^3  -40 q^5 + 8  q^7\right] , \nonumber
\\
\tilde B (q)&=  -6\mathrm{\mathrm{erfc}}(q) + \tpix \left[ 
-6q - 4q^3 + 32 q^5-8 q^7 \right]     \nonumber
\end{align}

\subsection{Flow from the $2a$ mode}
The mode $2a$ is usually called the rotlet.  In index notation, the flow due to the $2a$ mode is 
$$v_i^{2a}(\mathbf{x},\mathbf{x}_\alpha)
=- \tfrac12\epsilon_{jlm} \nabla_l G_{im} T^e_{\alpha,j}.$$
The velocity for $2a$ mode is:
\begin{align*}
    \left[v_{\alpha}^{2a}\right]^{\mathbb R}_i &=
     \frac{  D(\xi r)}{8\pi\eta}   \frac{\epsilon_{ijk} r_{k}}{2r^3}\,T_{\alpha,j}^e \\
         \left[v_{\alpha}^{2a}\right]^{\mathbb F}_i &= 
    \frac{C(\tfrac{k}{\xi})}{8\pi\eta}\sin(\mathbf k \cdot \mathbf r)\frac{\epsilon_{ijm}k_m  T^e_{\alpha,j}}{2k^2}
\end{align*}
Here we have defined:
\begin{align}
        {D}(q) &=   2{\mathrm{erfc}}(q)
 + \frac{2e^{-q^2}}{\sqrt{\pi}}\Big[2q -12 q^3  +4q^5\Big].
\end{align}
${C(\tfrac{k}{\xi})}$ is defined in Eq.\eqref{eq:ABC}
\subsection{Flow from the $2s$ mode}\label{sec:flow2s}
The mode $2s$ is usually called the stresslet \cite{grahamMicrohydrodynamics2018}. 
In index notation, the flow due to the $2s$ mode is 
$$v_i^{2s}(\mathbf{x},\mathbf{x}_\alpha)
= -c_2^s \nabla_l G_{im} V^{(2s)}_{\alpha,jm}.
$$
The velocity components are:
 \begin{align*}
     \left[v_{\alpha}^{2s}\right]^{\mathbb R}_i & = 
     \frac{-c^s_2}{8\pi\eta}
     \Big[
     { E_1 (\xi r)}\frac{\delta_{ik}r_j}{r^3}    
     +E_2 \frac{r_jr_k}{r^5}r_i
     \Big] V^{(2s)}_{\alpha,jk}
      \nonumber\\
            \left[v_{\alpha}^{2s}\right]^{\mathbb F}_i &=\frac{ c^s_2}{8\pi\eta}
 C(\tfrac{k}{\xi})\sin(\mathbf{k} \cdot \mathbf{r})
 \frac{\left(\delta_{ik}k_j - \hat{k}_j\hat{k}_kk_i\right)}{k^2}
 V^{(2s)}_{\alpha,jk}
 \end{align*}
\begin{align}
         E_1(q) &=   \frac{2e^{-q^2}}{\sqrt{\pi}}\Big[8 q^3 -4q^5\Big]\\
        E_2(q) &=  -3\mathrm{erfc}(q) - \frac{2e^{-q^2}}{\sqrt{\pi}}\Big[
        -3q + 2q^3 + 4q^5\Big]
\end{align}


\subsection{Flow from the $3s$ mode}\label{sec:flow3s}
In index notation, the flow due to the $3s$ mode is 
$$v_i^{3s}(\mathbf{x},\mathbf{x}_\alpha)
= c_3^s \nabla_l\nabla_m G_{ij} V^{(3s)}_{\alpha,jlm}.
$$
The real space velocity components are given by:
\begin{align*}
    \left[v_{\alpha}^{3s}\right]_i^{\mathbb R} &=\frac{c^s_{3}}{8 \pi \eta}\left[ \mathcal {U}(\xi r)\frac{\delta_{il} r_j r_k}{r^5} +\tilde{\mathcal {U}}(\xi r)\frac{r_l r_jr_k}{r^7}r_i\right]V^{(3s)}_{\alpha,l jk}\\
        \left[v_\alpha^{3s}\right]_i^{\mathbb F} &= 
c_3^s k_l k_m \hat{G}^{\mathbb{F}}_{ij} V^{(3s)}_{\alpha,jlm}
\end{align*}
Here, we have defined:\begin{align*}
\mathcal {U} (q)
         &=  -3\mathrm{erf}(q) +\tpix\big[-3q -2q^3 -20q^5 + 8\xi^7r^{7}\big] \nonumber\\
\tilde{\mathcal {U}}(q)&= 15\mathrm{erf}(q) +\tpix\Big(15q + 10q^3 + 4q^5 - 8\xi^7r^{7}\Big) 
\end{align*}
\subsection{Flow from the $3a$ mode}\label{sec:flow3a}
In index notation, the flow due to the $3a$ mode is 
$$v_i^{3a}(\mathbf{x},\mathbf{x}_\alpha)= c_3^a \nabla_{p}\left(\epsilon_{lmj}\nabla_m G_{ij} \right) V^{(3a)}_{\alpha,lp}.
$$
The  velocity components of the $3a$ mode as 
\begin{align*}
    \left[v_{\alpha}^{3a}\right]^{\mathbb R}_i &= c_3^a\frac{{S(\xi r)}}{8 \pi \eta}\frac{\epsilon_{ijk} r_j V^{(3a)}_{\alpha,km}r_m }{r^5}\\
    \left[v^{3a}_\alpha\right]^{\mathbb F}_i 
&= \frac{c_3^a}{8\pi\eta}\,C(\tfrac{k}{\xi}) \cos(\mathbf{k}\cdot \mathbf{r}) \epsilon_{ijm} \frac{k_{j} k_{n }}{k^2}  \,
V^{(3a)}_{\alpha,mn}
\end{align*}
Here,
\begin{align*}
\tilde{S}(q) = -6\mathrm{\mathrm{erfc}}(q)
 + \frac{2e^{-q^2}}{\sqrt{\pi}}\Big[  -16 q^3  + 32 q^5 - 8q^7\Big]
\end{align*}
\subsection{Flow for the $3t$ mode }
In index notation, the flow due to the $3t$ mode (also called a source dipole) is 
$$v_i^{3t}(\mathbf{x},\mathbf{x}_\alpha)= c_3^t \nabla^2  G_{ij}   V^{(3t)}_{\alpha,j}.
$$
The velocity components for the real part of the flow field are given as
\begin{align*}
    \left[v_\alpha^{3t}\right]^{\mathbb R}_i&=\frac{ c^t_3}{8\pi\eta}
    \left[\tilde A(\xi r)\frac{\delta_{ij}}{r^{3}} + {\tilde B(\xi r)}
    \frac{{r}_i{r}_j }{r^{5}}\right]
    V^{(3t)}_{\alpha,j}\\
        \left[v^{3t}_\alpha\right]^{\mathbb F}_i&=-\frac{ c^t_3}{8\pi\eta}{C(\tfrac{k}{\xi})}\cos(\mathbf{k}\cdot \mathbf{r}) \left(\delta_{ij}  - \hat {k}_i{\hat {k}}_j \right)V^{(3t)}_{\alpha,j}
\end{align*}
$\tilde A(q)$ and $\tilde B(q)$ are defined in appendix \ref{sec:flow1s}.
\subsection{Flow from the $4a$ mode of the slip}
\label{FLOW:4a}
In index notation, the flow due to the $4a$ mode is 
$$v_{i}^{4a}(\mathbf{x},\mathbf{x}_\alpha)= -c_4^a \nabla_{p}\nabla_{q}\left(\epsilon_{lmj}\nabla_m G_{ij} \right) V^{(4a)}_{\alpha,lpq}.
$$
The components of the flow in real space are given by
\begin{align*}
[{v_\alpha^{4a}}]_i^{\mathbb R} &= - 
   \frac{c^a_{4}}{8\pi\eta r^7}\Upsilon(\xi r)\Big( \epsilon_{ijk} r_j  r_m r_n  \Big)\,
   V^{(4a)}_{\alpha,kmn}
   \\
    \left[v^{4a}_\alpha\right]^{\mathbb F}_i &=   \frac{c^a_{4}}{8\pi\eta} 
    C(\tfrac{k}{\xi})\,\sin(\mathbf{k}\cdot \mathbf{r})
    \epsilon_{i jk} \frac{k_{j}k_{m} k_{n }}{k^2}
    V^{(4a)}_{\alpha,kmn}
    \nonumber 
\end{align*}
where the $\Upsilon$ given by
\begin{align*}
\Upsilon &= 30\mathrm{\mathrm{erfc}}(q)
 + \frac{2e^{-q^2}}{\sqrt{\pi}}\Big[ 6q  +32 q^3 - 32 q^5 - 80q^7 + 16\xi^9  r^9\Big] \nonumber \\
\end{align*}
\section{Irreducible components of RBM}
\label{app:RBM}
In this section, using Eq.\eqref{eq:mobMat} and \eqref{eq:PropTensors}, 
we write the explicit form of $\boldsymbol U^{l\sigma}_{\alpha\beta}$, and $\boldsymbol W^{l\sigma}_{\alpha\beta}$.
Expressions are obtained in terms of distance between source and field points: 
${r}^{\alpha\beta}=|\mathbf{r}^{\alpha\beta}|$, while 
$\mathbf{r}^{\alpha\beta} = \mathbf{x}_\alpha-\mathbf{x}_\beta$, 
where $\mathbf{x}_\alpha$ is the position of the $\alpha$-th particle and $\mathbf{x}_\beta$ is the position of the $\beta$-th particle. In this section, for brevity,  
use the notation that $\mathbf{r}=\mathbf{r}^{\alpha\beta}$. We also use ${\boldsymbol{\nabla}}=\boldsymbol{\nabla}_{\alpha}$ and 
$\tilde{\boldsymbol{\nabla}}=\boldsymbol{\nabla}_{\beta}$ for succinct expression in this section. 

\subsection{Rigid body motion from the $1s$ mode}\label{app:1sRBM}
In index notation, it is: 
\begin{align*}
U^{1s}_{\alpha\beta,i} =
\mathcal{F}^0\mathcal{F}^0\,G_{ij}F^e_{\beta,j},\quad 
W^{1s}_{\alpha\beta, i}=\tfrac12\epsilon_{ipm}\nabla_{p}\,G_{mj} F^e_{\beta,j}
\end{align*}
The explicit forms are:
\begin{align*}
       U^{1s}_{\alpha\beta,i}
       &=    \frac{F^e_{\beta,i}}{ 8\pi\eta}
     \left(A(\xi {r}) \frac{\delta_{ij}}{{d}} + B(\xi {r})\frac{{r}_{i} {r}_{j}}{r^3}\right)  
     \nonumber\\
     &+    \frac{F^e_{\beta,i}}{ 8\pi\eta}
    \frac{b^2}{3} 
          \left(\tilde A(\xi {r}) \frac{\delta_{ij}}{r^3} + \tilde B(\xi r)\frac{r_{i} r_j}{r^5}\right)  \\
  &+       F^e_{\beta,i} \left(  1 - \frac{b^2k^2}{3}  \right) \hat G^{\mathbb F}_{ij}(\mathbf{k})
       \\
  W^{1s}_{\alpha\beta,i}&=  \frac{\epsilon_{ijm}}{16\pi\eta}
  \bigg[{  D(\xi r)}    \frac{r_{m}}{r^3} 
  +C(\tfrac{k}{\xi})\sin(\mathbf k \cdot \mathbf r)\frac{k_m   }{k^2}
  \bigg] 
  F^e_{\beta,j}
  \end{align*}
See Eq.\eqref{eq:GrGf} for the definition of $\hat G^{\mathbb F}_{ij}$, Eq.\eqref{eq:ABC}  for $A(q)$ and $B(q)$, while $\tilde A(q)$ and $\tilde B(q)$ are in section \ref{sec:flow1s}.

\subsection{Rigid body motion from the $2a$ mode}
In index notation, it is: 
\begin{align*}
U^{2a}_{\alpha\beta, i}&=    - \tfrac12\epsilon_{jlm} \nabla_l G_{im} T^e_{\beta,j},
\quad
W^{2a}_{\alpha\beta, i }=-\tfrac14   \nabla^2\,G_{ij}   T^e_{\beta,j}
\end{align*}
The explicit form are:
\begin{align*}
U^{2a}_{\alpha\beta,i}&=  
\epsilon_{ijk}\bigg[\frac{  D(\xi r)}{8\pi\eta}   \frac{r_{k}}{2r^3} 
+  \frac{k_l G^{\mathbb F}_{kl} \tan (\mathbf{k} \cdot \mathbf{r})}{2}
  \bigg]   
  T^e_{\beta,j},\\
  W^{2a}_{\alpha\beta,i }&=  \frac{-1}{32\pi\eta} \left[\tilde A(\xi r)\frac{\delta_{ij}}{ r^3} +\tilde B(\xi r)\frac{ r_ir_j}{r^5} \right]  T^e_{\beta,j}\\
  &+ \frac{k^2}{4} \hat G^{\mathbb F}_{ij}(\mathbf{k})T^e_{\beta,j}
\end{align*}
\subsection{Rigid body motion from the $2s$ mode}
In index notation, the formal expressions are:
\begin{align}
      U^{2s}_{\alpha\beta,i}&=  c_2^s\mathcal{F}^0\mathcal{F}^1\,\tilde\nabla_{m}G_{ij}
      V^{(2s)}_{\beta,jm},\\  
       W^{2s}_{\alpha\beta,i}&=   \tfrac{1}{2}c_2^s 
\epsilon_{iu w}\nabla_{{u}}\tilde\nabla_{m}G_{w j}
V^{(2s)}_{\beta,jm},
\end{align}
The explicit forms are:
\begin{align*}
   U^{2s}_{\alpha\beta,i} 
    &=-
  \frac{c^s_2}{8\pi\eta}
  \Bigg[   \left(
      \frac{E_1}{r^3}(\xi r) \delta_{im}r_j 
+
     \frac{E_2}{r^5}r_ir_jr_m 
\right)
\\&- C(\tfrac{k}{\xi})\sin(\mathbf{k} \cdot \mathbf{r})
\left(
\frac{\delta_{ik}k_j}{k^2} - \frac{k_jk_m k_i}{k^4}\right)
\Bigg]V^{(2s)}_{\beta,jm}
\\
   W^{2s}_{\alpha\beta,i} 
     &=\frac{c^s_2}{16\pi\eta}
     \bigg[
\tilde{S}(\xi r)
     \frac{\epsilon_{ijk}{r_j r_m}}{r^3}
  \,V^{(2s)}_{\beta,jm}   \\
     &-
    C(\tfrac{k}{\xi})
\cos(\mathbf{k}\cdot \mathbf{r})\, \epsilon_{ipm}
\frac{k_{p} k_{j}}{k^2}  \, 
V^{(2s)}_{\beta,jm}
\bigg]
     \end{align*}
 $\tilde A(q)$ and $\tilde B(q)$ were defined in section \ref{sec:flow1s},
 $\tilde S$ is defined in section \ref{sec:flow3a}, 
 while ${E_1}$ , ${ E_2}$ are defined in section \ref{sec:flow2s}.
\subsection{Rigid body motion from the $3s$ mode}
In index notation, the formal expressions are:
\begin{align}
       U^{3s}_{\alpha\beta,i}
       &=c_3^s\mathcal{F}^0\mathcal{F}^2\,
\tilde\nabla_m
\tilde\nabla_p
G_{ij}  V^{(3s)}_{\beta,jmp},
\\
      W^{3s}_{\alpha\beta,i} 
      &= \tfrac{1}{2}c_3^s 
\epsilon_{it w} \nabla_{t}
\tilde\nabla_{m}
\tilde\nabla_{p}
G_{w j}  V^{(3s)}_{\beta,jmp}
\end{align}
The explicit forms are:\begin{align*}
   U^{3s}_{\alpha\beta,i}  
   &=\frac{c^s_{3}}{8 \pi \eta}\left[ \mathcal{U}\frac{ r_j r_m}{r^5}V^{(3s)}_{ijm} +\tilde{\mathcal {U}}\frac {r_j r_m r_p}{r^7}r_i V^{(3s)}_{\beta,jmp}\right] 
   \\&+ \frac{c^s_{3}}{8 \pi \eta}
   \left[\mathrm V^{(3s)}_{\beta,ijk} \frac{ k_j k_m}{k^2}+ V^{(3s)}_{\beta,jmp} 
   \frac{k_j k_m k_p}{k^4} k_i\right]\\
       W^{3s}_{\alpha\beta,i}&=
       \frac{c^s_{3}}{8 \pi \eta}\Big[ 
          \frac{\Upsilon(\xi r)}{r^7}\Big( \epsilon_{ijk} r_j  r_m r_n  \Big)\,
   V^{(3s)}_{\beta,kmn}
   \\
  &+    C(\tfrac{k}{\xi})\,\sin(\mathbf{k}\cdot \mathbf{r})
    \epsilon_{i jk} \frac{k_{j}k_{m} k_{n }}{k^2}
    V^{(4a)}_{\alpha,kmn}       \Big]
\end{align*}
$\mathcal{U}$ and $\title{\mathcal{U}}$ are defined in \ref{sec:flow3s}. Here, we have defined:
\begin{align}
     {\mathcal{U}'}(q) &=   3{\mathrm{erfc}}(q)
 + \frac{2e^{-q^2}}{\sqrt{\pi}}\Big[ 12q -32 10 q^3  -16 q^7\Big] , 
 \label{eq:S2}
 \end{align}
\subsection{Rigid body motion from the $3t$ mode}\label{app:3tRBM}
In index notation, the formal expressions are:
\begin{align}
     U^{3t}_ {\alpha\beta,i}  
     = c^t_3 \, {\nabla}^2{ {G}}_{ij}  V^{(3t)}_{\beta,j},\qquad 
       W^{3t}_ {\alpha,i}  =0.
\end{align}
Explicit forms are:
\begin{align*}
   U^{3t}_{\alpha\beta,i} 
   =
    {c^t_{3}}\left[
    \frac{\tilde A(\xi r)}{8\pi\eta}
    \frac{\delta_{ij}}{r^3} + 
        \frac{\tilde B(\xi r)}{8\pi\eta}
\frac{ r_ir_j}{r^5}-k^2 \hat{G}^\mathbb {F}_{ij}(\mathbf{k})\right]V^{(3t)}_{\beta,j}
\end{align*} 
 $\tilde A(q)$ and $\tilde B(q)$ were defined in section \ref{sec:flow1s} and  Eq.\eqref{eq:GrGf} gives the definition of $\hat G^{\mathbb F}_{ij}$
\subsection{Rigid body motion from the $3a$ mode}
In index notation, the formal expressions are:
\begin{align}
      U^{3a}_{\alpha\beta,i}&=  c_3^a \tilde\nabla_{p}\left(\epsilon_{lmj}\tilde\nabla_m G_{ij} \right) V^{(3a)}_{\beta ,lp},\qquad 
        \\W^{3a}_{\alpha\beta,i}&= \tfrac{c_3^a}{2}\epsilon_{ijk}\nabla_j \tilde\nabla_{p}\left(\epsilon_{lmt}\tilde\nabla_m G_{it} \right) 
 V^{(3a)}_{\beta, lp}
\end{align}
Explicit forms are:
\begin{align*}
U^{3a}_{\alpha\beta,i}&
=  \frac{c_3^a\,\tilde{S}(\xi r)}{8 \pi \eta}\bigg[\frac{\epsilon_{ijk} r_j r_m }{r^5}\bigg]V^{(3a)}_{\beta,km} \\
&
+
\frac{c_3^a}{8\pi\eta}\,C(\tfrac{k}{\xi})
\cos(\mathbf{k}\cdot \mathbf{r})\, \epsilon_{ipm} \frac{k_{p} k_{j}}{k^2}  \, V^{(3a)}_{\beta,jm}
\\
 W^{3a}_{\alpha\beta,i} &
 = \frac{c^a_{3}}{16\pi\eta}
 \Bigg[
\left(
\frac{48\tilde S+8S'}{3r^5} \delta_{ij}{r_m} -
\frac{5\tilde S-S'}{3r^7}{{{{r_i}} {r_j}} {r_m}}
 \right)
  \\&
 +  C(\tfrac{k}{\xi}) \frac{\sin(\mathbf{k}\cdot\mathbf{r})}{k^2 }\Big(  {k}^2~\delta_{ij}{k_m } - {k_ik_jk_m} \Big)   
 \Bigg]
V^{(3a)}_{\beta,jm}
\end{align*}
where $\tilde S$ is defined in section \ref{sec:flow3a} and
\begin{align*}
  S'&=  
  \frac{2e^{-q^2}}{\sqrt \pi}\Bigg[6q  -48q^3  + 192 q^{5} - 120q^{7} + 16q^{9}\Bigg]
\end{align*}
Eq.\eqref{eq:ABC} gives the definition of $C(q)$.
\subsection{Rigid body motion from the $4a$ mode}
In index notation, the formal expressions are:
\begin{align}
      U^{4a}_{\alpha\beta,i}&=  c_4^a \tilde\nabla_{p}\tilde\nabla_{u}\left(\epsilon_{lmj}\tilde\nabla_m G_{ij} \right) V^{(4a)}_{\beta, lpu},\qquad 
        \\W^{4a}_{\alpha\beta,i}&= \tfrac{c_3^a}{2}\epsilon_{ijk}\nabla_j \tilde\nabla_{p}\tilde\nabla_{u}\left(\epsilon_{lmt}\tilde\nabla_m G_{it} \right) 
 V^{(3a)}_{\beta, lpu}
\end{align}
Explicit forms are:
\begin{align*}
  U^{4a}_{\alpha\beta,i}&=  \frac{c^a_{4}}{8\pi\eta} \bigg[
   - \frac{\Upsilon(\xi r)}{r^7}\Big( \epsilon_{ijk} r_j  r_m r_n  \Big)\,
   \\
 &+  
    C(\tfrac{k}{\xi})\,\sin(\mathbf{k}\cdot \mathbf{r})
    \epsilon_{i jk} \frac{k_{j}k_{m} k_{n}} {k^2}
 \bigg]\,V^{(4a)}_{\beta,jmp} 
 \\
    W^{4a}_{\alpha\beta,i}
    &=\frac{c^a_{4}}{16\pi\eta}\bigg[ \frac{6}{r^{5}}\Big[{\mathcal{E}}_1\,
    {{r_i}{r_j}}{{r_m}}{{r_p}} - {\mathcal{E}}_2\,\delta_{ij} {r_m}{{r_p}}
  \Big]\\
  &+ C(\tfrac{k}{\xi}) \cos(\mathbf{k}\cdot\mathbf{r})
   (k^2\delta_{ij}  -  {{k_i}k_j})\frac{k_mk_p}{k^4}  \bigg]\,V^{(4a)}_{\beta,jmp} 
\end{align*}
Here, we have defined
$${\mathcal{E}}_1= 7\Upsilon +\tilde{\mathcal P},\qquad 
{\mathcal{E}}_2= 3\Upsilon -\tilde{\mathcal P}.$$ 
See \ref{FLOW:4a} for definition of $\Upsilon$, while 
\begin{align*}
\tilde{\mathcal P} &= 
\frac{2e^{-q^2}}{\sqrt{\pi}}\Big[ -24q + 84 q^3   + 160 q^5 -336q^7 \\
&+ 304 q^9  - 32q^{11}  \Big].
\end{align*}

\section{Extra contributions for an absolutely convergent flow expression}\label{app:convergence}
The expression of fluid flow in Eq.\eqref{eq:BIE-PBC} is 
rendered absolutely convergent by considering contributions from an imaginary surface $\Gamma'$ of length
scale much larger than the colloidal size \cite{glendinning1982pairwise, brady1988dynamic, ishikawa2008development} such that it contains finite number of particles. 
Thus, we need to consider contributions to the flow from the surface $\Gamma'$, which are:
\begin{alignat}{1}
 \boldsymbol v^{\mathrm{E}}=
\int\big[\mathbf{G}\cdot\boldsymbol{\sigma}+\mathbf{K}\cdot\boldsymbol{v}\big]\cdot\mathbf{n}\,\mathrm{d}{\mathcal S}_{\Gamma'}\label{eq:convFlow}
\end{alignat}
{ 
Note that this surface $\Gamma'$ always remains in the fluid, 
it is not allowed to cut any particle. 
Thus, the surface $\Gamma'$, has variations at
length scales of the order of the size of the colloids. 
To tackle the variations on the surface, we compute the 
integral on a new surface
$\Gamma$ which is allowed to cut the colloids, and thus, has no variations
at the colloidal scale \cite{glendinning1982pairwise}. 
}
Quantities in the integral of Eq.\eqref{eq:convFlow}
can then be replace by the respective suspension averages $\langle\dots\rangle$
\cite{batchelor1970stress}. 
Thus. Eq.\eqref{eq:convFlow} becomes:
\begin{alignat}{1}
 \boldsymbol v^{\mathrm{E}}&= \int\big[\mathbf{G}\cdot\langle\boldsymbol{\sigma}\rangle+\mathbf{K}\cdot\langle\boldsymbol{v}\rangle+\boldsymbol{\nabla}\mathbf{G}\cdot\langle\mathbf{Q}\rangle\big]\cdot\mathbf{n}\,\mathrm{d}{\mathcal S}_\Gamma
\end{alignat}
Here, $\mathbf{Q}$
is the additional contribution due to the change to surface $\Gamma$ \cite{glendinning1982pairwise, brady1988dynamic, ishikawa2008development}. It is: 
$$\mathbf{Q}= \frac{3}{8\pi b}\int\mathbf{Y}^{(2)}(\hat{\boldsymbol b})\,\mathbf{f}\,dS+\frac{\mathbf{I}}{8\pi b}\int\mathbf{f}\,dS.
$$
Finally, the volume enclosed by the surface $\Gamma$ is assumed to be infinite. 
The regularization procedure generates
extra contributions, which are then combined with Eq.\eqref{eq:BIE-PBC} to obtain an absolutely convergent expression of fluid flow \cite{glendinning1982pairwise, brady1988dynamic, ishikawa2008development}. The above contributions can be computed numerically to obtain convergent flow expression \cite{glendinning1982pairwise, brady1988dynamic, ishikawa2008development}.
It is useful to note that the flow contributions from terms $l>3$ are already convergent, while those till $l=3$ are normalized from the above procedure, such that we get an absolutely convergent expression for the fluid flow. 

%

\end{document}